\documentclass[onecolumn]{article}

\usepackage{graphicx}
\usepackage{authblk}

\begin{document}

\title{Steering efficiency of a ultrarelativistic proton beam in a thin bent crystal}

\author[1]{E. Bagli} 
\author[1]{L. Bandiera}
\author[1]{V. Guidi}
\author[1]{A. Mazzolari}
\author[2]{D. De Salvador}
\author[3]{A. Berra}
\author[3]{D. Lietti}
\author[3]{M. Prest}
\author[4]{E. Vallazza}
   
\affil[1]{INFN Sezione di Ferrara, Dipartimento di Fisica e Scienze della Terra, Universit{\`a} di Ferrara Via Saragat 1, 44122 Ferrara, Italy}
\affil[2]{ INFN Laboratori Nazionali di Legnaro, Viale dell'Universit{\`a} 2, 35020 Legnaro, Italy $\&$ Dipartimento di Fisica, Universit{\`a} di Padova, Via Marzolo 8, 35131 Padova, Italy }
\affil[3]{ Universit{\`a} dell'Insubria, via Valleggio 11, 22100 Como, Italy $\&$ INFN Sezione di Milano Bicocca, Piazza della Scienza 3, 20126 Milano, Italy}
\affil[4]{ INFN Sezione di Trieste, Via Valerio 2, 34127 Trieste, Italy}

\date{\today}

\maketitle

\begin{abstract}
Crystals with small thickness along the beam exhibit top performance for steering particle beams through planar channeling. For such crystals, the effect of nuclear dechanneling plays an important role because it affects their efficiency. We addressed the problem through experimental work carried out with 400 GeV/c protons at fixed-target facilities of CERN-SPS. The dependence of efficiency vs. curvature radius has been  investigated and compared favourably to the results of modeling. A realistic estimate of the performance of a crystal designed for LHC energy including nuclear dechanneling has been achieved.
\end{abstract}


\section{Introduction}
High-energy particles interacting with a properly oriented crystal can be captured within neighboring atomic planes and travel through the crystalline structure, i.e. particles do undergo channeling \cite{Dansk.Fys.34.14}. By proper shaping and bending of the crystals, channeling can be adopted to efficiently manipulate relativistic positive and negative particle beams \cite{Tsyganov682,Tsyganov684} as well as for e.m. radiation production \cite{RevModPhys.40.611,doi:10.1142/S0217751X10049876,1748-0221-3-02-P02005,PhysRevA.79.012903,PhysRevA.86.042903,Lietti,Bandiera}. Bent crystals have already been proposed \cite{MokhovSSCL} and demonstrated for beam collimation at Tevatron \cite{ipac10carrigan}, SPS \cite{Scandale201078} and U-70 \cite{U70collimation} accelerators. In addition, bent crystals have also been adopted for beam steering \cite{Elishev1979387} and extraction \cite{BellaziniReportCrystal,Akbari1993491,PhysRevLett.87.094802,PhysRevSTAB.5.043501,FlillerIII200547} in circular accelerators, as well as for splitting and focusing of external beams \cite{Denisov1992382}. Moreover, collimation and extraction of TeV-proton and ions with bent crystals have been proposed as upgrades of LHC \cite{Scandale:1357606,Rakotozafindrabe:1492806,Lansberg:1503414}.

\begin{figure}[ht]
\includegraphics[width=3.4in]{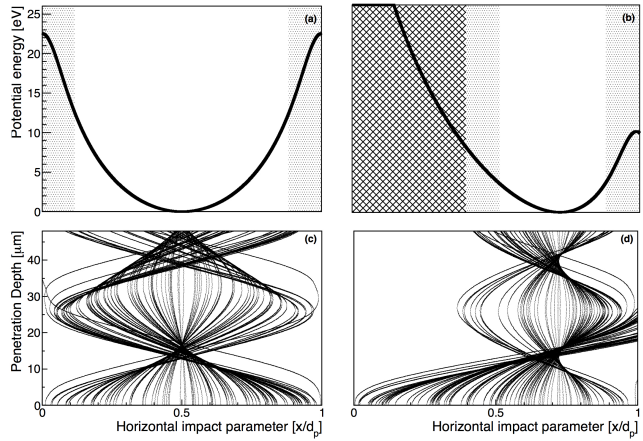}
\caption{\label{fig1} (a)-(b) Potential energy in the reference frame integral to crystal planes for $R = \infty$ (a) and $R = 3R_{c}$ evaluated through the ECHARM software \cite{PhysRevE.81.026708}. (b). Gray boxes highlight impact region which will cause particle to dechannel due to multiple scattering on nuclei and core electrons. Crossed box region is the impact region of particle with do not channel because of the presence of the crystal curvature. (c)-(d) Trajectories of 400 GeV/c proton interacting with a Si crystal with bending radius $R = \infty$ (c) and $R = 3R_{c}$ (d).}
\end{figure}

The key parameter regarding utilization of bent crystals for beam manipulation is deflection efficiency, i.e. the fraction of particles captured into the channeling state for the whole of the crystal. As some examples, performance of beam extraction and collimation via channeling is critically linked to the fraction of the beam deflected \cite{Scandale201078,Lansberg:1503414}. Channeled particles may suffer dechanneling as a consequence of multiple scattering with either valence electrons or with nuclei and core electrons \cite{RevModPhys.46.129}, namely via electronic and nuclear dechannelings, respectively \cite{PhysRevB.8.3117,PhysRevB.38.4404,Taratin1990247,taratin98}. Nuclear dechanneling occurs for particles impinging close to the atomic planes (see Fig. \ref{fig1}). Such particles traverse a short distance in the crystal channel before they are dechanneled. On the other hand, electronic dechanneling holds for the particles far from the atomic planes (see Fig. \ref{fig1}). Such particles slowly increase their transverse energy via interaction with the electrons until they reach the region with high-atomic density where they are rapidly dechanneled. The rates of nuclear and electronic dechanneling are quite different. As an example, for $400$ GeV/c protons interacting with Si (011) planes, $L_{e}\sim220$ mm \cite{Carrigan}, while $L_{n}\sim1.5$ mm \cite{Scandale2009129},  $L_{e}$ and $L_{n}$ being the electronic and nuclear dechanneling lengths, respectively. For a bent crystal, channeling efficiency is ulteriorly spoiled. In fact, the depth and the width of the potential well are smaller than for a straight crystal due to the centrifugal force acting on channeled particles (see Fig. \ref{fig1}.b).

The advent of a new generation of crystals manufactured through micro machining techniques allowed fabricating crystals with an unprecedented small thickness along the beam, which is comparable to or even much lower than the nuclear dechanneling length. Thus, usage of such crystals allowed to measure nuclear dechanneling length and to record efficiencies larger than the maximum level foreseen \cite{Scandale2009129}. "Thin" crystals in circular accelerators boosted extraction efficiency  \cite{U70collimation} and demonstated the possibility to be used as primary collimators \cite{Scandale2012231}.

Such crystals pointed out the necessity to revise the physical models to describe the dynamics of channeling features. In fact, most of previous theories were suited for the available "thick" crystals, i.e. for crystals with length along the beam much longer than the nuclear dechanneling length \cite{Sun198460,Biryukov1994245,Maller1994434,Biryukov}.

In this article we report a systematic study of the channeling efficiency for a thin bent crystal vs its bending radius. A campaign of experimental measurements and three different theoretical approaches to calculate channeling efficiency in thin bent crystals are here presented and compared.

\section{Experimental}
The experiment was performed at external line H8 of the Super Proton Synchrotron (SPS). A single strip-like crystal offering (110)  planes for channeling was shaped to the size of 1 x 55 x 1.96 mm$^{3}$. The crystal was fabricated by means of silicon anisotropic etching techniques \cite{0022-3727-41-24-245501,baricordi:061908}. The strip was mechanically bent through a purposely fabricated crystal holder made of Al \cite{ipac10tors}. Primary curvature along the 55 mm size was imposed through the holder to cause a secondary curvature driven by anticlastic deformation \cite{AfoninU70JETP,guidi:113534} on the 1.96 mm size. The holder was mounted on a high-resolution two-axis goniometer with accuracy of 1.0 $\mu$rad on both movements. The crystal was exposed to a beam of $400$ GeV/c protons with $10.2\pm0.1$ $\mu$rad horizontal RMS divergence and $8.0\pm0.1$ $\mu$rad vertical RMS divergence. Particle trajectories were tracked before and after interaction with the crystal thanks to a telescope made by three double-sided Si microstrip detectors \cite{PhysRevLett.101.234801}.

The bent strip was aligned to attain the condition for planar channeling far from alignment with main [111] crystal direction to avoid interference by axial channeling. Optical pre-alignment of the strips without the beam was accomplished through a laser system, which allowed evaluating the relative angle between the strip face parallel to the beam and the beam direction. After that, an angular scan with the goniometer was performed in order to determine the best channeling alignment through the evaluation of the efficiency of deflection of the strips. When the maximum efficiency had been recorded, a low-statistic run was performed to measure the crystal torsion. A precise screw-system installed on the strip-holder was adopted to compensate for strip torsion induced by mechanical stresses \cite{ipac10tors}. The effect of residual torsion on channeling efficiency was corrected through a specific selection algorithm to add horizontal angular shift to an incoming particle proportional to its vertical impact parameter. Evaluation of strip torsion was first performed by studying the dependence of the maximum of efficiency peak on the horizontal incoming angle and vertical position. Precise measurement of residual torsion are reported in Tab. \ref{tab1}. Five curvatures of the crystal were considered and the whole procedure was repeated for each of them.

Data analysis was performed over a 800 x 2000 $\mu$m$^2$ portion of the incoming beam centered on the strip. Analysis of channeling deflection efficiency was done by selecting a 2 $\mu$rad wide region of horizontal angle of incoming particles over the observed peak of the maximum of channeling efficiency. The distribution of the outgoing horizontal deflection angle was fitted with one gaussian for the channeling peak, one for the undeflected peak and an exponential for the fraction of dechanneled particles between the two peaks, i.e., for the fraction of the particles not channeled at full bending angle. Channeling deflection efficiency was computed as described in Ref. \cite{PhysRevSTAB.11.063501}. Measured efficiencies for the five bending radii are reported in Tab. \ref{tab1} and shown in Fig. \ref{fig3}.

\section{Modeling}
Three models have been worked out to describe channeling efficiency in thin bent crystals. The first model is based on statistical and geometrical considerations (analytical), the second on the solution of the equation of motion with simplified surrounding conditions (semi-analytical) and the third on full Monte Carlo simulation of the particle trajectories in the crystal (Monte Carlo). Because of the increasing calculation complexity, the computational time changes by orders of magnitudes between the first and the third methods. Therefore, each of the models can be useful depending on the purpose for which an evaluation of channeling efficiency is made.

\subsection{Analytical}
The first model has been developed extending the approach developed for thick crystals in Refs. \cite{Biryukov,Baurichter200027}. By considering a parallel beam of protons interacting with a thick straight crystal, particles undergoing nuclear dechanneling ($N_n$) are immediately dechanneled and efficiency falls off. The remaining fraction of the channeled particles are subject to electronic dechanneling, which can be described by diffusion theory \cite{Biryukov} and approximated through an exponential decay function with decay length equal to $L_{e}$.  For  slightly bent crystals the harmonic approximation well describes the interplanar potential far from atomic planes. The width of the potential well is reduced by a quantity proportional to the decrease in the width of the well. In addition, the electronic dechanneling length scales as a function of $R$, because particle trajectories closely approach atomic nuclei in a bent crystal (see Fig. \ref{fig1}). Under harmonic approximation the scaling factor is $(1-R_{c}/R)^2$ \cite{Biryukov}, which is equal to the lowering of the potential well depth. From previous considerations, channeling efficiency in thick bent crystals $\epsilon_{L}$ holds

\begin{equation}
\epsilon_{l}(R,L)=\frac{N_{tot} - N_n}{N_{tot}}(1-R_{c}/R)^2e^{ -\frac {L} {(1-R_{c}/R)^2L_{e}}}
\label{eq1}
\end{equation}

where $L$ is the crystal length along the beam direction, $N_{n}$ is the number of particles which can be subject to nuclear dechanneling and $N_{tot}$ is the total number of particles interacting with the crystal.

For a thin bent crystal, the same approximation can be adopted through mild modifications. Since the crystal thickness is so thin to be comparable to nuclear dechanneling length, not all the $N_{n}$ particles have abandoned the channeling state at the crystal exit. Therefore, Eq. \ref{eq1} for deflection efficiency has to be modified to include the possible contribution of the fraction of particles that close-encounter nuclei at the entrance of the crystal. As a result, the channeling efficiency for thin bent crystal $\epsilon_{s}(R,L)$ holds \cite{Scandale201370}

\begin{equation}
\epsilon_{s}(R,L)=\frac{N_n}{N_{tot}}e^{ -\frac {L} {L_{n}}}+\epsilon_{l}(R,L)
\label{eq2}
\end{equation}

\subsection{Semi-analytical}
Instead of a simple geometric model, a more detailed description of the dechanneling process can be achieved by studying the dechanneling probability as a function of the intensity of interaction for particle at given impact parameter. Scatterings with electrons and nuclei does not significantly alter the trajectory of a channeled ultra-relativistic particle. Such approximation can be adopted to treat the crystals with length considerably longer than one oscillation period $\lambda$, i.e., for the overwhelming majority of practical cases. In the model, the transverse energy variation is computed only after one oscillation period. Such energy variation depends on the intensity of interaction of channeled particles with nuclei and electrons, which is in turn a function of the average quantity of matter $\overline{\rho}$ encountered by a channeled particle during its motion. Since no interaction is considered during particle motion within one oscillation, the particle trajectory is a function only of the initial transverse energy $E_{T,0}$, which is related to the impact parameter $x_{0}$ and the incoming angle $\theta_{0}$.

\begin{equation}
E_{T,0}=\frac{pv}{2}\theta_{0}^2 + U(x_{0})
\label{eqTrEnergyStraight}
\end{equation}

and is evaluated through the integration of the relativistic equation of motion

\begin{equation}
x(z) = \frac{1}{\lambda}\int_{\lambda}  \sqrt{\frac{2}{pv}\left[E_{T,0}-U(x)\right]  }  {\mathrm{d}z}
\label{eqTrajStraight}
\end{equation}

$x$ being the particle position in the coordinate orthogonal to crystal planes, $z$ the direction along the particle motion, $p$ and $v$ the particle momentum and velocity, $U(x)$ the interplanar potential. Examples of integration are shown in Fig. \ref{fig1} (c).
The nuclear density averaged over one period $\overline{\rho}(x_0)$ holds

\begin{equation}
\overline{\rho}_n(x_0)=\frac{1}{\lambda}\int_{\lambda}\rho_n(x(z)){\mathrm{d}z}
\label{eqRHO}
\end{equation}

where ${\rho}_n(x)$ is the distribution of nuclear density averaged over planes or axes. Considering the electronic density ${\rho}_e(x)$, the same equation keeps true for the electron density averaged over planes or axes, $\overline{\rho}_e(x_0)$. The probability of dechanneling is equivalent to the probability to receive enough transverse energy to be kicked out of the potential well. Transverse energy can be acquired through multiple scattering with nuclei or loss of energy by collisions with electrons. Indeed any variation of the total energy modifies particle direction with respect to the orientation of crystal plane. Consequently, the kinetic transverse energy may vary.  Therefore, the probability to be dechanneled holds

\begin{equation}
\frac{\mathrm{d}P(x_0)}{\mathrm{d}x_0} =  \frac{1}{d_p}{\int}_{{{\Delta}E}_{x}(x_0)}^{+\infty}   {\frac{\mathrm{d}P_{E_x}(\overline{\rho})}{\mathrm{d}E_x}} {\mathrm{d}E_{x}}
\label{eqEN}
\end{equation}

where $ {{\mathrm{d}P_{E_x}(\overline{\rho})}/{\mathrm{d}E_x}}$ is the distribution function of energy acquired by the particle after one oscillation and ${\Delta}E_{x}(x_0)$ is the energy to overcome the potential well, i.e., for dechanneling. As the crystal is bent, the equations for the particle's trajectory and for the transverse energy variation needed for dechanneling have to be changed to take into account the lowering of the potential barrier. In particular, Eq. \ref{eqTrEnergyStraight} changes to

\begin{equation}
E_{T,0}=\frac{pv}{2}\theta_{0}^2 + U(x_{0})+\frac{pvx_{0}}{R}
\end{equation}

As a result, Eq. \ref{eqTrajStraight} becomes

\begin{equation}
x(z) = \frac{1}{\lambda}\int_{\lambda}  \sqrt{\frac{2}{pv}\left[E_{T,0}-U(x)-\frac{pvx}{R}\right] } {\mathrm{d}z}
\label{eqTrajBent}
\end{equation}

Eq. \ref{eqTrajBent} is quite general being valid for any shape of the potential $U(x)$. Then, Eqs. \ref{eqRHO} and \ref{eqEN} can be repeated for calculation of dechanneling probability. Examples of integration with Moli{\`e}re potential are shown in Fig. \ref{fig1}.d. The calculation of the averaged potential, electron density and nuclear densities of the crystal have been worked out though the ECHARM software \cite{PhysRevE.81.026708}, which allows to evaluating the averaged electrical characteristics of a complex atomic structure and to choose between various models for the form factor of the electron density. Then, the trajectory's equation is numerically solved. The efficiency is computed as the fraction of particles which reach the crystal end without dechanneling.

\subsection{Monte Carlo}
The third method is the most accurate though the most time-consuming because it calculates each particle's trajectory by solving the equation of the motion. To date, coherent interaction of particle with crystal has been prevalently aided by Monte Carlo codes. Binary collision model \cite{FLUX7} and continuum potential approximation \cite{taratin98} were the approaches mostly adopted to simulate the interaction. Recently, a new model has also been developed which solves the equation of the motion in three dimensions as for binary collision models in a volume following the particle in its motion \cite{Sushko2013404}. Since the continuum potential approximation proved to reproduce the experimental results at high-energies with high accuracy, the DYNECHARM++ Monte Carlo code \cite{Bagli2013124} based on such approximation has been adopted. In the simulation, Eq. \ref{eqTrajBent} is numerically integrated as for the previous model, but the total transverse energy may vary at each step, not only at the end of one oscillation period. Averaged electrical characteristics have been evaluated through the ECHARM software \cite{PhysRevE.81.026708}. Therefore, the particle is being tracked along the whole crystal length. Interaction with nuclei and electrons within the channel are taken into consideration according to Ref. \cite{PhysRevB.8.3117}. The efficiency is evaluated as in the semi-analytical case.

\section{Discussion and conclusions}
In order to compare the models with the experimental results, the same input parameters have been used for the calculations. The fraction of particles impinging onto the nuclear corridor is $\sim19.5\%$ \cite{Scandale201370}, the critical radius and the oscillation period are $R_c\sim0.7$ m and $\lambda\sim67.5{\mu}m$ \cite{Biryukov}, respectively. Approximation of experimental form factor of Ref. \cite{Su:sp0113} has been adopted to describe the electron density for a Si atom for the analytical calculation and Monte Carlo simulation. Calculated efficiencies for the three models are reported in Tab. \ref{tab2} and visually superimposed to the experimental results in Fig. \ref{fig3}.

As clearly shown in Fig. \ref{fig3}, the three models exhibit rather similar trends. The prevision of theoretical estimates lie within $\pm5\%$ in efficiency from the corresponding experimental values. Though, the Monte Carlo method outputs the most accurate values. The dependence on radius for all the three models well follows the experimental curve, except for the zone near the critical radius. The discrepancy in this case has to be ascribed to the lack of knowledge of the exact density distribution between atomic planes. Indeed, dechanneling probability significantly changes with the atomic distribution in the structure because of the different average density of nuclei and electrons encountered  by a particle in its trajectory. In particular, the more the crystal is bent the more the particles are pushed against the high-density atomic region, i.e., the zone within which nuclear dechanneling takes place. To study the dependence of efficiency on atomic density, we have adopted two different models for the atomic form factor, i.e, the Moli{\'e}re approximation and the approximation based on the experimental x-ray diffraction data. Channeling efficiency between the calculation of the above form factors differs by $+0.4\%$ and  $+3.0\%$ at $R/R_c$ equal to $40.6$ and $3.3$, respectively. Since, the efficiencies do not scale proportionally with the bending radius, the need for more precise experimental measurement of nuclei and electronic density arises to better evaluate and simulate channeling efficiency in a thin highly bent crystal.

The experimental data and the calculation models presented in this paper can be used also to predict the channeling performance scaling at various energy by the parametrization of crystal geometrical characteristics. Channeling efficiency for thin crystals has been shown to depend mainly on two factors, the length $L$ and the bending radius $R$ of the crystal. Thus, an efficiency surface $\epsilon_{ch}(R,L)$ can be built for any energies by varying such factors. Since the critical radius and the nuclear dechanneling length  depend on the particle energy, one can expresses $R$ in unit of  $R_{c}$ and $L$ in unit of $L_{n}$ and the same $\epsilon_{ch}(R,L)$ is valid for all the relativistic energy range for positive particles. The efficiency surface is shown in Fig. \ref{fig4}.

The efficiency surface is a fast way to evaluate the most performing geometrical features for a bent crystal. Recently proposed upgrades of the LHC have highlighted the possibility to use bent crystal for beam manipulation. The fabrication of a  bent crystal is strictly connected to three factors, i.e., the damage yield to radiation, the behavior with highly charged ions and the deflection efficiency \cite{Uggerhj200531}. Experiments with Si strips either under single-pass \cite{Scandale20102655} or multi-pass \cite{Altuna1995671,Elsener1996215,Scandale201078} channeling have shown that coherent interaction is a high-efficiency process that strongly reduces the total number of nuclear interactions with respect to interaction with an amorphous material for protons and high-Z ions \cite{Biino2002417,Uggerhoj2005240,Scandale2011547,Scandale2012231}.

The dependence on the particle energy of the critical radius is well known in the literature \cite{Biryukov}, $R_{c}(E){\propto}E$. On the contrary, the analytical dependence of $L_{n}$ on the energy has not been studied yet. In order to attain such trend, the semi analytical method has been adopted. Interaction of protons with various energies impinging on a straight Si crystal aligned along (110) planes have been simulated. Rechanneling has not been considered in the simulation.  Fig. \ref{fig5} shows the dependence of channeling efficiency for a collimated beam on particle energy and crystal length. 

From Eq. \ref{eq1} and \ref{eq2} and bearing in mind that $L_{n}{\ll}L_{e}$, the inefficiency $1-\epsilon_{s}$ can be approximated by $\sim{L/L_{n}}(N_{n}/N_{tot})$. Thus, by fixing the inefficiency, ${L_{n}}$ is proportional to crystal length $L$. From previous consideration for Si (110) crystals, by setting $L=L_{n}$, we obtain that a crystal as thick as the dechanneling length provides a channeling efficiency $\sim87\%$ independently from the particle energy. By imposing $L\sim{E^{m}}$, contour curve corresponding to the $87\%$ efficiency level can be fitted by a second order curve, showing that $L_{n}$ is proportional to the square root of the beam energy (see Fig. \ref{fig5}).

In addition, the semi-analytical calculation method has been adopted to work out the efficiency for a Si (110) crystal with $0.1$ mrad fixed bending angle and variable length, as in Refs. \cite{Baurichter200027,Uggerhj200531}. By comparing results shown in Fig. \ref{fig6}.b with those in the literature, e.g. Refs. \cite{Baurichter200027,Uggerhj200531}, all the methods agree on the optimal length to be chosen for a crystal for LHC though they disagree on the reachable maximum efficiency with about $\sim10$ $\%$ discrepancy. In fact, for all the models the effect of curvature radius is taken into account in the same manner, but the influence of the dechanneling is evaluated differently. The simulation of the interaction between the LHC beam and a bent crystal should be performed paying great attention to the influence of dechanneling. As a consequence, a model with accurate description of the nuclear dechanneling in a thin crystal has to be adopted, instead of model which relies on theory well suited for a thick crystal. In the models of Ref. \cite{Baurichter200027,Uggerhj200531} particles which can be subject to nuclear dechanneling are regarded as immediately dechanneled by nuclei. However, though not mentioned explicitly, $N_{n}$ is arbitrarily set at 0. in Ref. \cite{Baurichter200027} and 0.1 in Ref. \cite{Uggerhj200531}. On the contrary, in the model of this paper such particles are not immediately dechanneled and $N_n$ is computed from the crystal characteristics.

In summary, in this article the channeling efficiency for a thin bent crystal has been experimental studied for a Si crystal. Three different models and their features have been discussed and compared to experimental data. The surface efficiency has been proposed to scale the experimental results at any relativistic energy. The evaluation of the best geometrical features of a bent crystal for the
The design and fabrication of suitable crystals for manipulation of high-energy beam such as the future upgrades of LHC.

We are grateful to Professor L. Lanceri (INFN and University of Trieste) who provided the tracking detectors. We acknowledge partial support by the INFN ICE-RAD and PRIN 2008TMS4ZB projects.

\bibliographystyle{ieeetr}
\bibliography{biblio}

\begin{table}[htdp]
\caption{\label{tab1} Experimentally measured parameters of channeling. Radius is the bending radius of the crystal, $\Delta\theta$ the mean horizontal deflection angle, $\tau$ the torsion of the strip, $\epsilon$ the deflection efficiency evaluated as in Ref. \cite{PhysRevSTAB.11.063501}}
\begin{center}
\begin{tabular}{ c || c | c | c }
Radius (m) & {$\Delta\theta$ ($\mu$rad)} & $\tau$ ($\mu$rad/mm) & {$\epsilon$ ($\%$)} \\ \hline
28.4$\pm$0.4 & 69$\pm$1 & 12$\pm$1 & 81$\pm$4\\  \hline
18.4$\pm$0.2 & 107$\pm$1 & -9$\pm$1 & 80$\pm$3\\  \hline
6.8$\pm$0.1 & 289$\pm$3 & 1$\pm$2 & 71$\pm$1\\  \hline
3.5$\pm$0.1 & 555$\pm$4 & 2$\pm$2 & 57$\pm$1\\  \hline
2.3$\pm$0.1 & 847$\pm$5 & 11$\pm$1 & 34$\pm$4
\end{tabular}
\end{center}
\end{table}%

\begin{table}[htdp]
\caption{\label{tab2} Channeling efficiency of experimental data (Exp.) and calculation with analytical (An.), semi analytical (S.a.) and Monte Carlo (MC) methods. Values are also shown in Fig. \ref{fig3}}
\begin{center}
\begin{tabular}{ c || c | c | c | c}
$R/R_c$ & Exp. & An. & S.a. & M.C\\ \hline
40.6  & 81 & 81.1 &  83.1 & 81.2\\  \hline
26.3 & 80 & 79.0 &  81.9 & 79.7\\  \hline
9.7 & 71 & 68.2 & 74.1 & 72.3\\  \hline
5.1 & 57 & 56.3 & 54.1 & 56.8\\  \hline
3.3 & 34 & 43.7 & 40.42 & 39.9
\end{tabular}
\end{center}
\end{table}%

\begin{figure}[ht]
\includegraphics[width=3.4in]{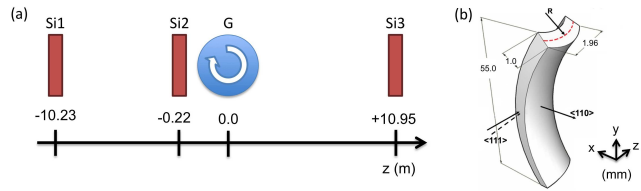}
\caption{\label{fig2} (a) Experimental setup used during data taking at H8 external line of the SPS accelerator at CERN. Three Si strip detector (S) were mounted $\sim10$ m one from another in order to reach a $\mu$rad angular resolution in the reconstruction of the particle track. The crystal was mounted on a high-resolution two-axis (G) offering an accuracy of 1.0 $\mu$rad. (b) A sketch of the bent Si strip.}
\end{figure}

\begin{figure}[ht]
\includegraphics[width=3.4in]{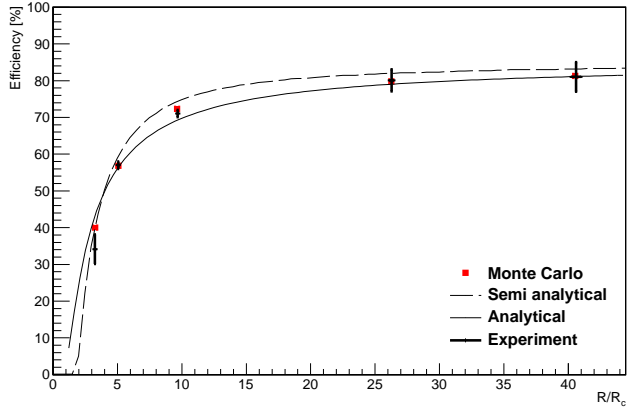}
\caption{\label{fig3} Channeling efficiency plotted Vs bending radius ($R$) over critical radius ($R_c$) for experiments with 400 GeV/c protons impinging on a Si crystal and for the analytical, the semi analytical and the Monte Carlo calculation methods.}
\end{figure}

\begin{figure}[ht]
\includegraphics[width=3.4in]{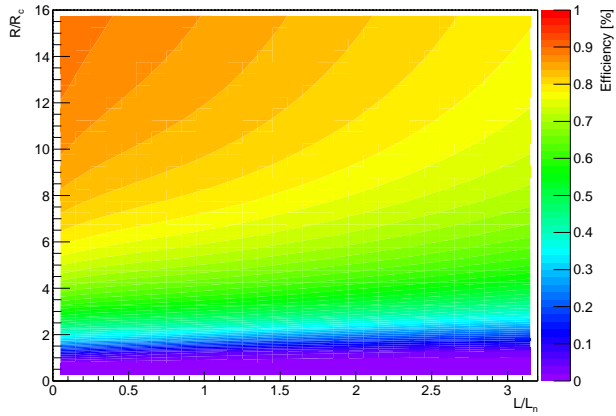}
\caption{\label{fig4} Efficiency vs. radius ($R$) over critical radius ($R_{c}$) and length ($L$) over nuclear dechanneling length ($L_{n}$) for Si (110) strip exposed to a collimated proton beam computed with the semi analytical method.}
\end{figure}

\begin{figure}[ht]
\includegraphics[width=3.4in]{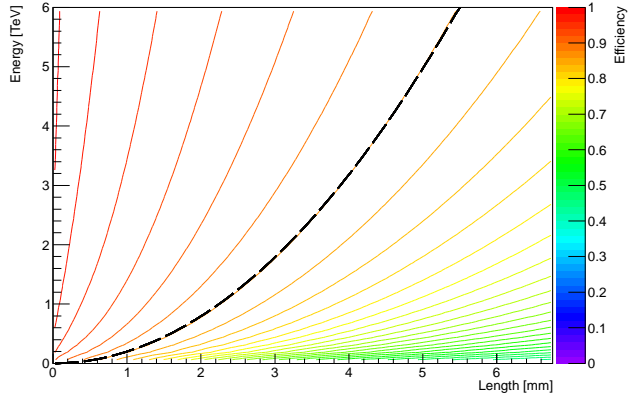}
\caption{\label{fig5} Efficiency vs. energy ($E$) and length ($L$) for Si (110) strip exposed to a collimated proton beam computed with the semi analytical method. Dashed line is a second order curve which fit the graph for efficiency equal to $87\%$.}
\end{figure}

\begin{figure}[ht]
\includegraphics[width=3.4in]{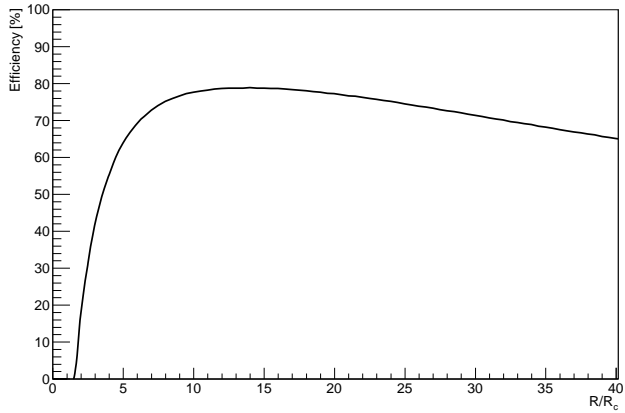}
\caption{\label{fig6} Efficiency vs. $R/R_{c}$ for $7$ TeV proton beam impinging on a Si (110) strip with fixed $0.1$ mrad bending angle computed with the semi analytical method.}
\end{figure}

\end{document}